\documentclass[twocolumn]{aastex63}
\usepackage{times}
\usepackage{amsmath}
\usepackage{graphicx}
\usepackage{CJK}
\graphicspath{ {./images/} }

\newcommand{\beq}{\begin{equation}}
\newcommand{\eeq}{\end{equation}}

\begin{document}
\begin{CJK*}{UTF8}{gbsn}
%%%%%%%%%%%%%%%%%%%%%%%%%%%%%%%
\title{Radiative diffusion in a time-dependent outflow: a model for fast blue optical transients}

\author{CHUN CHEN (陈春)}
\affiliation{School of Physics and Astronomy, Sun Yat-Sen University, Zhuhai, 519082, China}
\affiliation{CSST Science Center for the Guangdong-Hongkong-Macau Greater Bay Area, Sun Yat-Sen University, Zhuhai, 519082, China}

\author{RONG-FENG SHEN (申荣锋)}
\affiliation{School of Physics and Astronomy, Sun Yat-Sen University, Zhuhai, 519082, China}
\affiliation{CSST Science Center for the Guangdong-Hongkong-Macau Greater Bay Area, Sun Yat-Sen University, Zhuhai, 519082, China}

%%%%%%%%%%%%%%%%%%%%%%%%%%%%%
\begin{abstract}
Fast Blue Optical Transients (FBOTs) are luminous transients with fast evolving (typically $t_{\rm rise}<12\  \rm days$) light curve and blue color (usually $\rm {-0.2\ >\ g-r\ >\ -0.3}$) that cannot be explained by a supernova-like explosion. We propose a radiative diffusion in a time-dependent outflow model to interpret such special transients. In this model, we assume a central engine ejects continuous outflow during a few days. We consider the ejection of the outflow to be time-dependent. The outflow is optically thick initially and photons are frozen in it. As the outflow expands over time, photons gradually escape, and our work is to model such an evolution. Numerical and analytical calculations are considered separately, and the results are consistent. We apply the model to three typical FBOTs: PS1-10bjp, ZTF18abukavn, and ATLAS19dqr. The modeling finds the total mass of the outflow ($\sim 1-5 {M_{\odot}}$), and the total time of the ejection ($\sim$ a few days) for them, leading us to speculate that they may be the result of the collapse of massive stars. 
\end{abstract}


%%%%%%%%%%%%%%%%%%%%%%%%%%%%%%
\section{Introduction}
As the observed data increases significantly \citep[e.g.,][]{Kaiser2002,Law2009,Brown2013,Shappee2014,Chambers2016,Kulkarni2018}, astronomical transients attract more and more attention in recent years. Fast Blue Optical Transients (FBOTs) are a type of transients, which show some features as follow: fast evolving (typically $t_{\rm rise}<12\  \rm days $ ), high bolometric luminosity ($L_{\rm bol} \sim 10^{43}\  \rm erg\  s^{-1}$), and blue color (usually $\rm {-0.2\ >\ g-r\ >\ -0.3}$) \citep[]{Coppejans2020,Drout2014,Pursiainen2018,Terasaki2020}. The observed temperature declines from 20000 - 30000 K to 10000 K within the initial 10 days  \citep[]{ Drout2014, Kuin2019}. Some FBOTs have been observed luminous X-ray emission and the delayed radio radiation including AT2018cow, CSS161010, AT2020xnd \citep[]{Margutti2019, Coppejans2020, Bright2021}.

These features are different from the other transients. No observed evidence proves that there is gamma ray in FBOTs, therefore, ruling out a gamma ray burst (GRB). Typical rise times for tidal disruption events (TDEs) could be weeks to months and the decay times could be longer \citep[]{Perley2019}, thus TDEs were unlikely to explain FBOTs. 

Radio emission of FBOTs \citep[]{Coppejans2020,Anna2020} suggests a possible existence outflow. For supernova, typically $t_{\rm decline} \sim$ {a few months} is much slower than FBOTs, and $\rm {g-r\ >\ -0.2}$ \citep[]{Drout2014} is much higher than FBOTs. Also ordinary supernova models are hard to explain these features. On one hand, $\sim$ a few $M_{\odot}$ $^{56}Ni$ should be needed to achieve such peak luminosity in supernova model, but according to the rise timescale, the total ejecta mass would be only 0.01${M_{\odot}}$, which is contradictory with the $^{56}Ni$ mass. On the other hand, the shock breakout model of a supernova could exhibit a fast-rising luminosity, but the typical rise time scale is shorter than $\rm{2-3}$ days \citep[]{Waxman2017}. If we apply the shock breakout model to FBOTs, the dimension of the progenitor should be $\rm{10^{14}}\  \rm{cm}$, which is comparable to that of a red supergiant. However, as a result of such a great massive envelope, the decline timescale would be slower. The light curve would show a plateau after the peak \citep[]{Perley2019,zheng2021}. 

Due to these special features of FBOTs, some work aim to propose novel mechanisms, including shock interaction with circumstellar medium \citep[e.g., ][]{Drout2014,Leung2019,Fox2019,Tolstov2019,Perley2019,Anna2020,Ho2020,Rest2019,Tanaka2016,McBrien2019,Margutti2019,Rivera Sandoval2018}, which was  systematically studied in \cite{Suzuki2020}, and the evolution of outflow \citep[]{kashiyama15,piro20,Uno2020}. For the latter explanation, the model in \cite{kashiyama15} did not consider the evolution of outflow material in the early stage of material ejection, and models in \cite{piro20} and \cite{Uno2020} assumed that the velocity of the ejected material is constant.

The progenitors of FBOTs is also a controversial issue. Proposed scenarios include stellar explosion \citep[]{Coppejans2020,Margutti2019}, tidal disruption events \citep[]{Perley2019,Kuin2019,Kremer2020}, common envelope jets \citep[]{Soker2019}, and merger events between a white dwarf and a neutron star or a black hole \citep[]{Gillanders2020}. We will discuss about the progenitors based on our results.

In this paper, we assume a time-dependent outflow which carries radiation with it. As the outflow expands, it becomes optically thin and the radiation in the outflow is released. To explain how such an outflow is produced, we adopt a core-collapse model for the formation of the black hole \citep[]{kashiyama15,Perley2021}. In fact, after a massive star explodes, a black hole accretion disk system is usually produced \citep[]{Antoni2021}. A black hole forms as the result of core collapse of a massive star. While the inner portion of the stellar envelope directly falls into the nascent black hole, the outer materials with sufficient specific angular momentum will fall back and form an accretion disk. The radiation pressure in the accretion disk increases due to the shock wave and the viscous effect, which drives the outflow in the form of disk wind \citep[]{kashiyama15}. The outflow we discussed in this paper does not depend on the mechanism by which the outflow is produced, the core-collapse mechanism we mentioned is the most likely mechanisms that produces the outflow, but we do not rule out other mechanisms such as a compact binary merger \citep[]{McCully2017}, or a stellar tidal disruption events \citep[]{Perley2019, Coppejans2020}.

\cite{kashiyama15} considered an instantaneous ejected outflow, while \cite{piro20} and \cite{Uno2020} considered a uniform speed of the outflow. We assume that the ejection of outflow is within a period of time. Since the outflow ejection lasts for a few days, it is natural that the ejection velocity varies slowly with time within this long period, and it may also fluctuate on smaller time scales. For the latter case, the collisions within the outflow would redistribute the momentum in a way such that faster shells move ahead and slower shells trail behind. So we conclude that it is reasonable to have a velocity distribution of the outflow. We assume that the velocity of particles is distributed in the range of $(v_{\rm min}, v_{\rm max})$ \citep[]{kashiyama15, Tsuna2021}. We describe the physical process briefly and perform analytical calculations. Assuming the internal energy of the outflow is comparable to its kinetic energy according to the Equipartition Theorem, we model the temperature evolution and the light curve.  

In section \ref{section_2}, we describe the time dependence of the outflow and the velocity shells. In section \ref{section_3}, we calculate our models including the temperature evolution and the bolometric light curve predictions. We apply the model to observed data of three FBOTs in section \ref{section_4}, including PS1-10bjb \citep[]{Drout2014}, ZTF18abukavn \citep[]{Leung2021}, and ATLAS19dqr \citep[]{ chen2020,Prentice2020,zheng2021}. Finally, we discuss some implications and caveats in section \ref{section_5} and summarize the results in section \ref{section_6}.

\section{THE TIME-DEPENDENT OUTFLOW} \label{section_2}

A mass outflow (also named as wind or ejecta) 	is the subject of many models for transients. Usually the outflow is assumed to be spherically symmetric, only for the ease of treatment. It was ejected from a central engine, and the ejection was active for some finite duration. There might be a velocity distribution within the outflow. In the following, we describe our treatment of the mass / density / velocity distribution within this outflow. It serves as a framework upon which the photon diffusion process within the outflow is further calculated. 

%%%%%%%%%%%%%%%%%%%%%%%%%%%%%
\subsection{The mass-velocity distribution and the shells}

\begin{figure}
\includegraphics[scale=0.4]{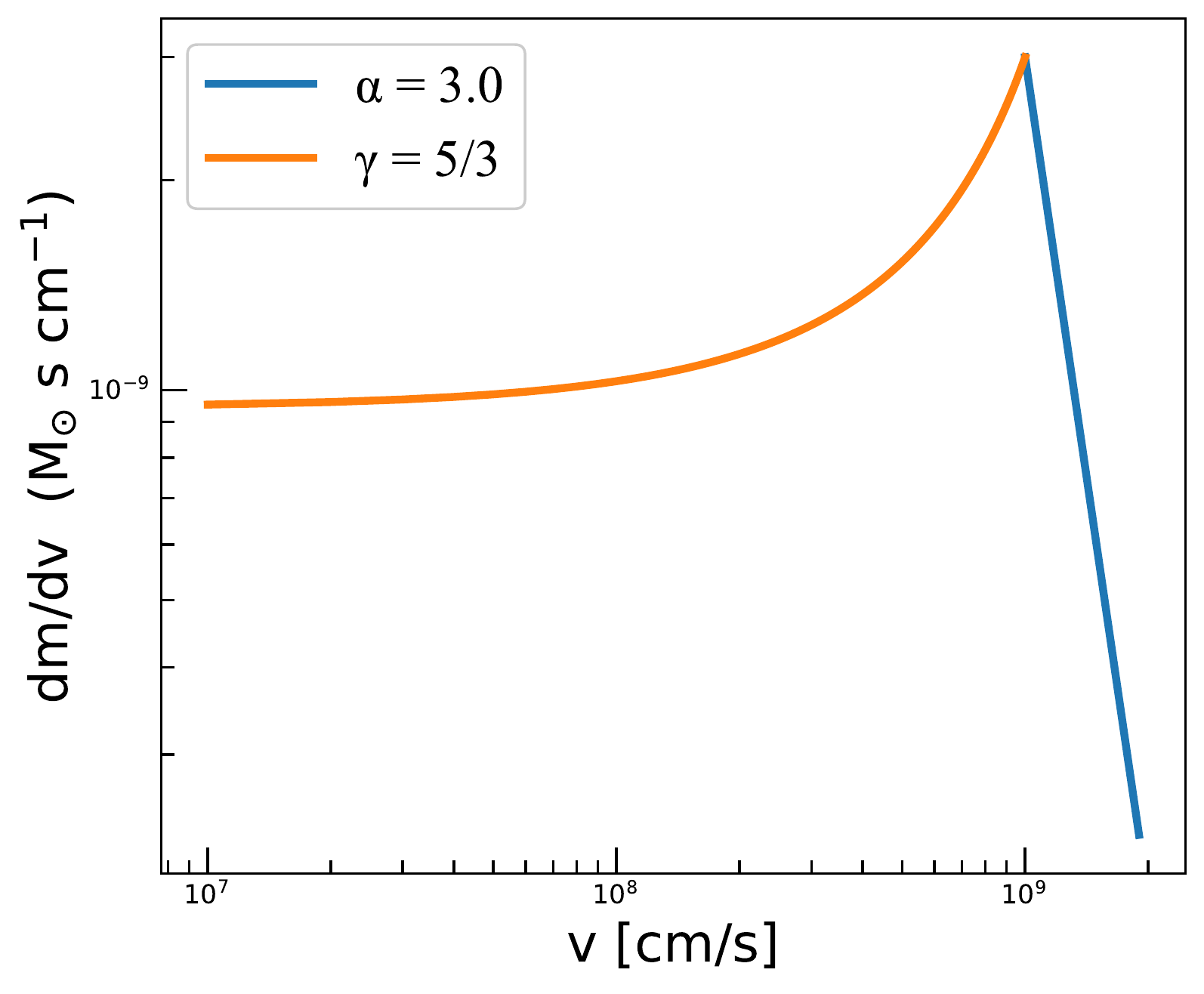}
\includegraphics[scale=0.4]{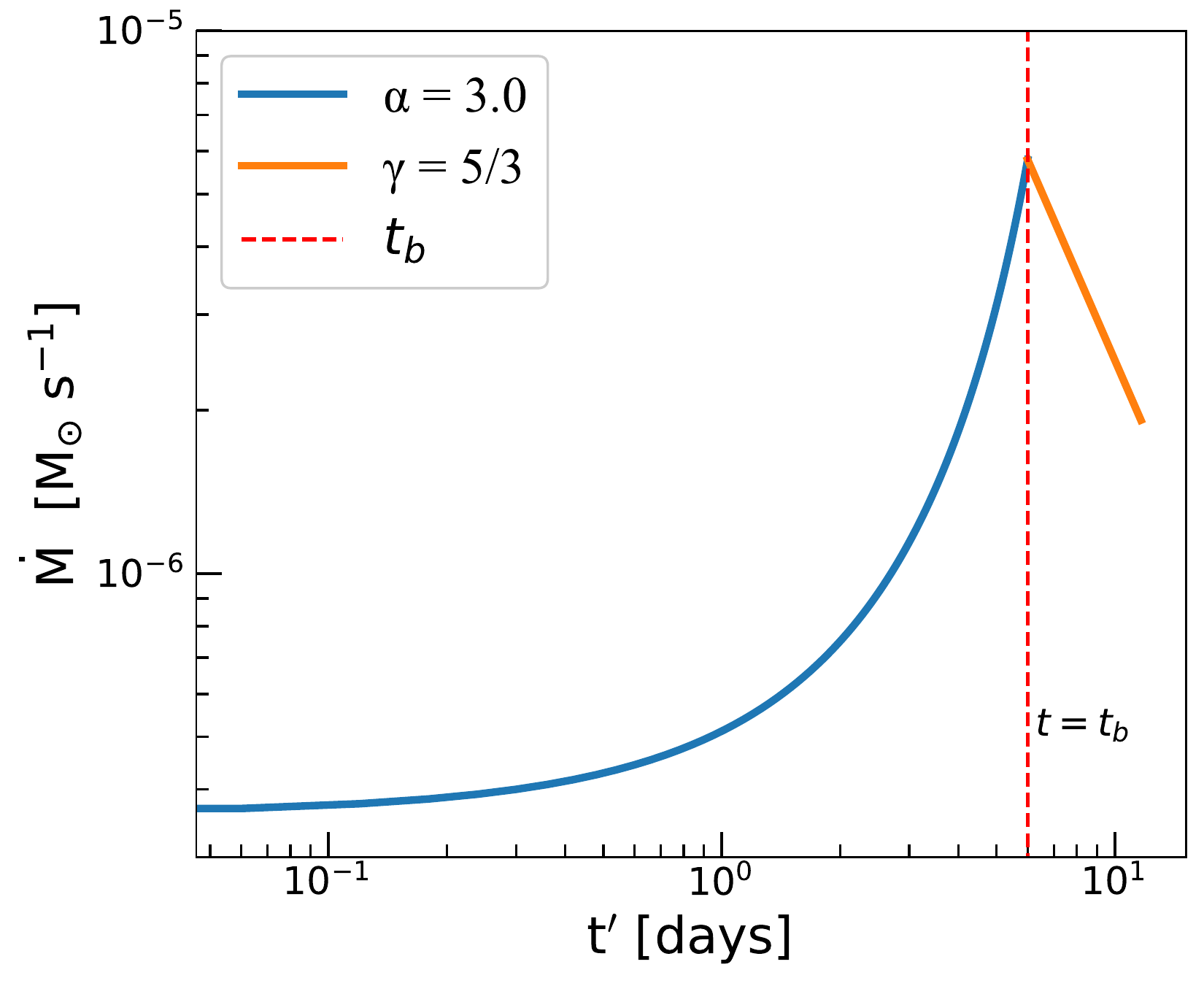}
\includegraphics[scale=0.4]{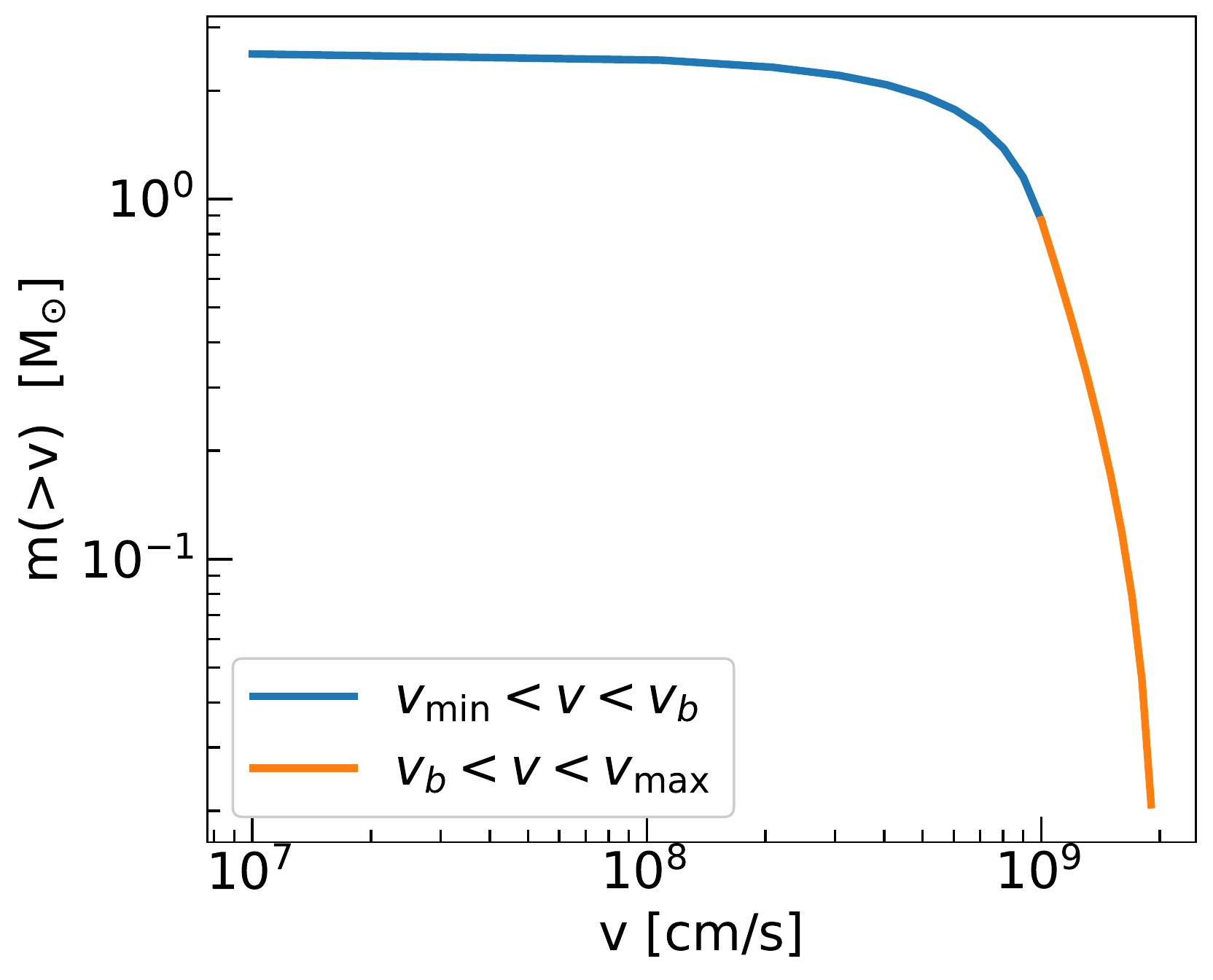} 
\caption{Top: The mass distribution of the outflow material in velocity space. Middle: Ejection rate evolves as time. Bottom: Total outflow mass as the function of inner velocity. We adopt $\alpha = 3.0$, $\gamma = 5/3$, $v_{\rm max}=0.067\  c $,  $v_{\rm b}=0.033\  c $, $v_{\rm min}=v_{\rm b}/100 $ and $M_{\rm out}=2.44 \  M_{\odot}$.}
\label{dm_dt_t}
\end{figure}

The ejection of the outflow lasts for a duration of $t_{\rm end}$, with a total mass of the ejecta $M_{\rm out}$. We divide the ejection into two stages: the rise and the decline of the outflow mass rate, respectively (see the top panel of Figure \ref{dm_dt_t}). We consider that the outflow is composed of a series of continuous mass shells. The terminal velocities of those shells are distributed in the range of $(v_{\rm min}, v_{\rm max})$ in a homologous manner, such that faster shells always move in front of the slower ones.

Assuming the mass distribution in the velocity space:
\beq \label{dm_dv}
\frac{dm}{dv} = \begin{cases}
 \alpha \frac{M_{1}}{v_{\rm b}}  \left(\frac{v}{v_{\rm b}}\right)^{-\alpha-1}, & \mbox{for} ~ v_{\rm b} < v < v_{\rm max}, \\
\alpha \frac{M_{1}}{v_{b}}  \left(\frac{v_{\rm max}-v}{v_{\rm max}-v_{b}}\right)^{-\gamma}, & \mbox{for} ~v_{\rm min} < v < v_{\rm b},
\end{cases}
\eeq 
where $v_b$ is the shell velocity that the shell ejected at the end of the ejection rate rising stage. $\alpha$ ($>0$) and $\gamma$ ($> 0$) are model parameters. A larger $\alpha$ ($>0$) means more mass is contained in slower shells while a larger $\gamma$ ($> 0$) means less mass is contained in slower shells. Since the total ejecta mass is $M_{\rm out}$, we can write $M_1$ as: 
\beq \label{M_1}
M_1=\frac{M_{\rm out}}{ 1+ \frac{ \alpha(v_{\rm max}-v_b)}{(1-\gamma)v_b} \cdot [(\frac{v_{\rm max}-v_{\rm min}}{v_{\rm max}-v_b})^{1-\gamma}-1]-(\frac{v_{\rm max}}{v_b})^{-\alpha}}.
\eeq

The mass ejection rate history $\dot{M}(t') \equiv dm/dt'= dm/dv \times (dv/dt')$ depends on the velocity-ejection time distribution, which we assume to be a simple linear relation:
\beq  \label{velocity-ejection time distribution}
v(t') = v_{\rm max} - (v_{\rm max} - v_{\rm b}) \frac{t'}{t_{\rm b}}.
\eeq
where $t_b$ is time that the slowest shell ejects during the ejection rate rising stage. Note that the fastest shell is ejected at $t'=0$, the slowest shell is ejected at $t_{\rm end}$, and it could be given as:
\beq
t_{\rm end}=\frac{(v_{\rm max}-v_{\rm min})}{(v_{\rm max}-v_{b})} t_{b}.
\eeq 

According to Eqs. (\ref{dm_dv}, \ref{velocity-ejection time distribution}), the mass ejection rate could be described as:
\beq \label{ejection rate 1}
\begin{split}
& \frac{dm}{dt^{\prime}}= \alpha \frac{(v_{\rm max}-v_b)}{v_{\rm b} t_b}  \left[\frac{v_{\rm max}t_b-(v_{\rm max}-v_b)t^{\prime}}{v_{\rm b} t_b}\right]^{-\alpha-1} \times \\
&\frac{M_{\rm out}}{1+\frac{\alpha(v_{\rm max}-v_b)}{v_b(1-\gamma)}[(\frac{v_{\rm max}-v_{\rm min}}{v_{\rm max}-v_b})^{1-\gamma}-1]-(\frac{v_{\rm max}}{v_b})^{-\alpha}}, \\& ~~ \mbox{for} ~ 0 < t^{\prime} < t_b,
\end{split}
\eeq

\beq  \label{ejection rate 2}
\begin{split}
&\frac{dm}{dt^{\prime}}= \alpha \frac{(v_{\rm max}-v_b)}{v_{\rm b} t_b}  \left(\frac{t^{\prime}}{ t_b}\right)^{-\gamma} \times \\ 
&\frac{M_{\rm out}}{1+\frac{\alpha(v_{\rm max}-v_b)}{v_b(1-\gamma)}[(\frac{v_{\rm max}-v_{\rm min}}{v_{\rm max}-v_b})^{1-\gamma}-1]-(\frac{v_{\rm max}}{v_b})^{-\alpha}}, \\
&~~ \mbox{for} ~ t_b <  t^{\prime} < t_{\rm end},
\end{split}
\eeq
where we adopt $\gamma=5/3$ in this paper. Since the envelop material will first fall back to form the disk with the fallback rate $\dot{M}_{\rm fb} \propto t^{-5/3}$  \citep[]{Michel1988, Zhang2008, Dexter2013}, the falling material will be immediately accreted by the black hole, and the mass of the accreted material is comparable to the mass of the ejected material \citep[]{kashiyama15},  we assume that the mass ejection rate is also $dm/dt^{\prime}  \propto t^{-5/3}$. 

Figure \ref{dm_dt_t} shows the evolution of the outflow mass rate with time, the differential and the integral mass distribution as a function of velocity, respctively.
%%%%%%%%%%%%%%%%%%%%%%%%%%%

\subsection{The density of the shells}
\begin{figure*}
\begin{center}
\includegraphics[scale=0.3]{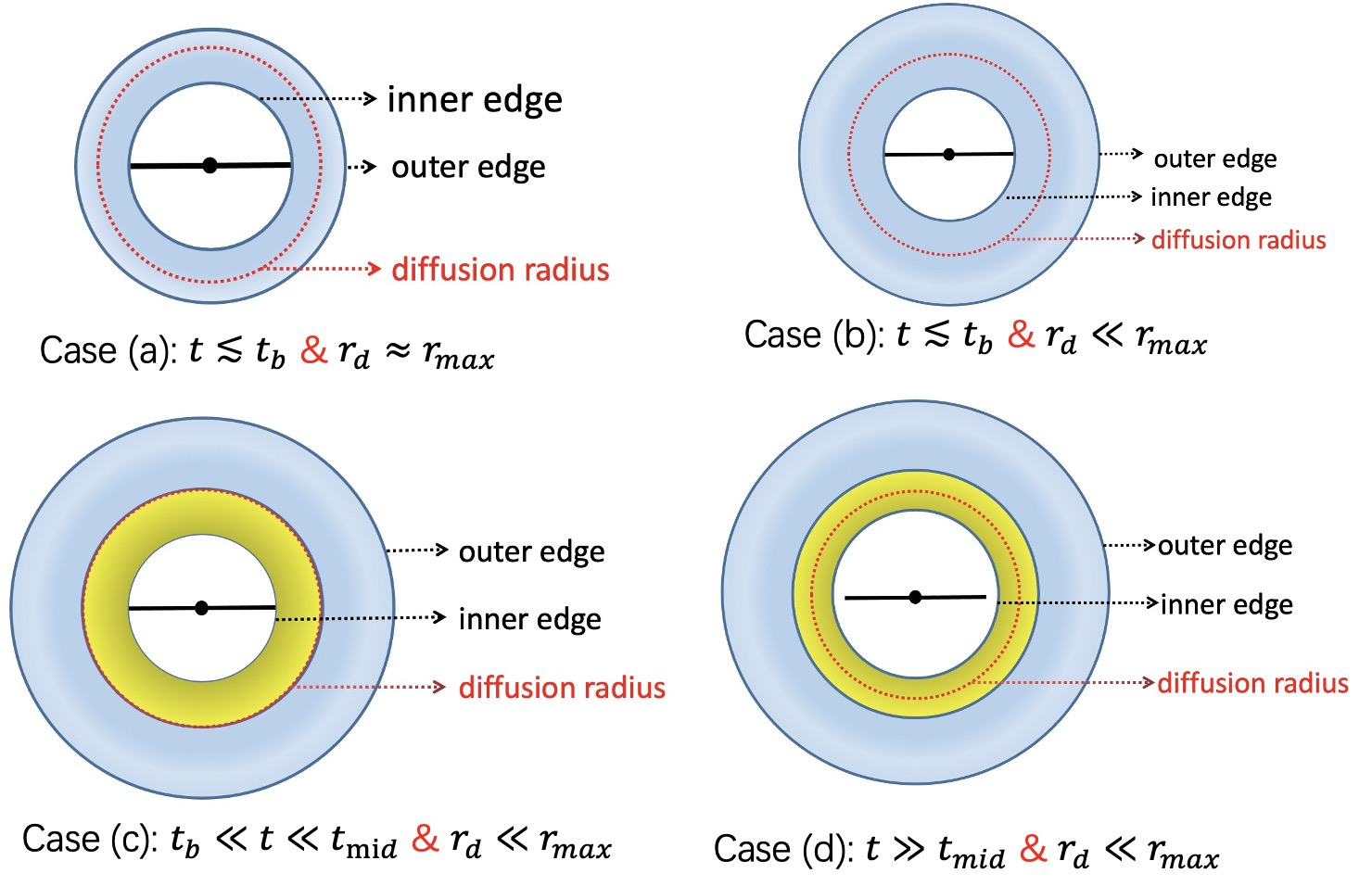}
\caption{The evolution of the diffusion radius $r_d$ during the whole process. At the center of each diagram is the black hole - accretion disk system where the outflow is ejected. The blue is the material that was ejected when the ejection rate went up, and the yellow is the material that was ejected when the ejection rate went down.}
\label{DF}
\end{center}
\end{figure*}

Consider one of those shells whose mass is $dm$ and its terminal velocity is $v$ -- we will call it the $m$-shell or $v$-shell hereafter. It was ejected at the time $t'$. At the current time $t$, the $m$-shell has moved to the radius
\beq
r(t)=r_0 + v (t-t'),
\eeq 
where $r_0$ is the launching radius. The width of the shell is 
\beq		\label{eq:dr}
d r= (t-t') d v - v dt',
\eeq
where the first term on the right-hand side accounts for the radial expansion due to the velocity difference, and the second term is the initial width. Since the velocity difference is $dv \approx (v_{\rm min} - v_{\rm max})dt'/t_b$ according to Eq. \eqref{velocity-ejection time distribution}, the shell width expression can be written in two asymptotic regimes:
\beq
dr \simeq \begin{cases}
- v dt', & \mbox{for} ~ (t-t') \lesssim t_b, \\
(t-t')dv, & \mbox{for} ~ (t-t') \gg t_b,
\end{cases}
\eeq 
for $(t-t') \lesssim t_b$, this case indicates that the width of the shell is dominated by the time difference between different velocity shells. And for $(t-t') \gg t_b$, this case indicates that the width is dominated by the velocity difference between different velocity shells.

The mass density of the shell is given by $dm= 4\pi r^2 \rho dr= (dm/dv)\times dv$. Thus:
\begin{eqnarray} \label{rho} 
\rho_v (t) & = & \frac{dm / d v}{4\pi (v (t-t')+r_0)^2} \cdot \frac{dv}{dr} \nonumber \\
 & = & \frac{dm / d v}{4\pi (v (t-t')+r_0)^2} \cdot \frac{1}{(t-t^{\prime})-v \frac{dt^{\prime}}{dv}} \nonumber \\
& = &  \frac{dm / d v}{4\pi (v (t-t')+r_0)^2} \nonumber \\
& \times & \begin{cases} 
 \left(\frac{(v_{\rm max}-v_{b})}{v t_{b}}\right), & \mbox{for}~~(t-t') \lesssim t_{b}, \\
 (t-t')^{-1}, & \mbox{for} ~~ (t-t') \gg t_{b}.  
 \end{cases}
\end{eqnarray}
where $r_0$ is the launch radius of the outflow. We consider this radius as 100 times the Schwarzschild radius. In this paper, we adopt $r_0 = 3\times 10^8 \rm cm$ according to the hypothesis that the central black hole is a stellar-mass black hole.

%%%%%%%%%%%%%%%%%%%%%%%%%%%%
\section{RADIATIVE PROPERTIES} \label{section_3}

%%%%%%%%%%%%%%%%%%%%%%%%%%%
\subsection{The internal energy and temperature}

The internal energy of the outflow is dominated by photons. Thus for an individual shell, the evolution of its energy density obeys the adiabatic law $aT^4 \propto \rho^{4/3}$. Let $T_0$, $\rho_0$ be the initial temperature and density, respectively, of the shell when it was ejected, at the initial radius $r_0$. Throughout the paper, we consider $r_0$ be much smaller than $r$. Assuming the following energy equipartition at the base between the internal and kinetic energies:
\beq \label{initial energy}
aT_0^4 = \frac{\eta}{2} \rho_0 v^2,
\eeq 
where $\eta$ is a constant, we have
\beq		\label{eq:T4}
a T^4 = \frac{\eta}{2} \rho v^2 \left(\frac{\rho}{\rho_0}\right)^{1/3},
\eeq
since $v$ is defined as the outflow terminal velocity, it can be considered as constant for $r \gg r_0$.

We can calculate the equivalent luminosity: for $t - t^{\prime} \lesssim t_b$, we could obtain $L(r)  = 4 \pi r^2 a T^4 v = \eta \dot{M}(r) v^2 (\rho / \rho_0)^{1/3} /2$ according to Eqs. \eqref{eq:T4}, and according to Eqs. \eqref{rho}, $(\frac{\rho}{\rho_0})^{1/3} \sim [\frac{v(t^{\prime} - t^{\prime}) +r_0}{v(t- t^{\prime}) +r_0}]^{2/3} \sim {(\frac{r_0}{r})^{2/3}}$. For $t - t^{\prime} \gg t_b$, we obtain $(\frac{\rho}{\rho_0})^{1/3} \sim [\frac{(v(t^{\prime}-t^{\prime})+r_0)^2vt_b}{(v(t-t^{\prime})+r_0)^2 (v_{\rm max}-v_{b}) (t-t^{\prime})}]^{1/3} \sim [\frac{r^2_0 v^2t_b}{(v_{\rm max}-v_{b})}]^{1/3}\frac{1}{r}$. To summarize the results, we can write as:

\begin{eqnarray}		\label{eq:Lcarry}
L(r) & = &4 \pi r^2 a T^4 v \nonumber \\
  & \sim & \frac{\eta}{2} \dot{M}(r) v^2  \times 
  \begin{cases} 
  \left(\frac{r_0}{r}\right)^{2/3}, & \mbox{for}~ t - t^{\prime} \lesssim t_b,\\
  (\frac{r^2_0 v^2t_b}{(v_{\rm max}-v_{b})})^{1/3} \frac{1}{r}, & \mbox{for}~ t - t^{\prime} \gg t_b
  \end{cases} 
\end{eqnarray}
for the photons that  are `frozen' in and carried by the outflow, up to a diffusion radius $r_d$ where they start to diffuse out.  Note that the outflow mass rate $\dot{M}(r) \equiv 4\pi r^2 \rho v$ is smaller than $\dot{M}(t')$, because of the radial expansion that the shell experienced. Eq. \eqref{eq:Lcarry} suggests that the radiative luminosity will always be a tiny fraction of the outflow's kinetic energy luminosity.

%%%%%%%%%%%%%%%%%%%%%%%%%%%%
\subsection{The photon diffusion and luminosity}

The optical depth for the $v$-shell is 
\beq
\tau_v \equiv \int_{r_v}^{r_{\rm max}} \kappa \rho dr =\int_{v_d}^{v_{\rm max}} \kappa \rho \frac{dr}{dv} dv,
\eeq
we adopt $\kappa =\  0.4 \ cm^{2}\ g^{-1}$ in this paper.

Using Eq. \ref{rho}, we have
\beq \label{eq:tau}
\tau_v =\int_{v_d}^{v_{\rm max}} \kappa \frac{dm / d v}{4\pi (v (t-t')+r_0)^2} \cdot \frac{dv}{dr} \frac{dr}{dv} dv,
\eeq
where $r_{\rm max}= v_{\rm max} t$ is the outer edge of the outflow. 
We adopt the following formula in \cite{piro20} to estimate the photon diffusion time of the $v$-shell
\beq
t_{\rm dif} \approx \tau_v \frac{r_v}{c} \frac{(r_{\rm max} -r_v)}{r_{\rm max}},
\eeq
which matches the expected limits: $t_{\rm dif} \approx \tau_v (r_{\rm max} -r_v)/c$ when $r_v \approx r_{\rm max}$, and $t_{\rm dif} \approx \tau_v r_v/c$ when $r_v \ll r_{\rm max}$. Note that the definition of $t_{\rm dif}$ in \cite{kashiyama15} corresponds to the first limit only.

The dynamical time of the $v$-shell is the current time $t \approx r_v/v$. The diffusion radius $r_d$ (or, the diffusion $v$-shell) is defined as the radius of the shell for which the two time scales are equal $t_{\rm dif} = t$. Thus:
\beq  \label{eq:difshell}
\begin{split}		
\tau_v \equiv \int_{r_d}^{r_{\rm max}} \kappa \rho dr = \int_{v_d}^{v_{\rm max}} \kappa \rho \frac{dr}{dv} dv \\
 =  \frac{r_{\rm max}}{(r_{\rm max} -r_d)} \frac{c}{v_d}.
\end{split}
\eeq

This shows that the classic condition $\tau \approx c/v$ typically used for defining the diffusion radius \citep[e.g.,][]{strubbe09,nakar10} is valid only when the diffusion radius is deep inside the outflow \citep{piro20}.

The observed luminosity is the energy flux of photons diffusing across the diffusion radius:
\beq		\label{eq:Ld}
L_{\rm obs} = 4\pi r_d^2 a T^4(r_d) \left(v_d - \frac{d r_d}{dt}\right),
\eeq 
where following \cite{piro20} we include the $d r_d / dt$ term to account for the changing position of the diffusion radius.

The diffusion radius evolves during the whole stage as in Figure \ref{DF}. We define $t_{\rm mid}$ as the time when the diffusion radius coincides with the radius of the $v_b$-shell, i. e., $r_d(t_{\rm mid})=v_b(t_{\rm mid}-t^{\prime})$.

For any time $t$, the diffusive shell (i.e., $v_d$, $r_d$) would be known by solving Eq. \eqref{eq:difshell}, then the observed color temperature $T_{\rm obs} \equiv T(r_d)$ and luminosity $L_{\rm obs}$ can be calculated.

%%%%%%%%%%%%%%%%%%%%%%%%%%%%%%%
\subsection{Temporal properties of the diffusive emission}
\begin{table*}
\begin{center}
\caption{Asymptotic scalings of the evolution of the observed color temperature and bolometric luminosity.}
\label{Analytical results}
\begin{tabular}{|l|l|l|l|l|}
\hline
Stages & $t \lesssim t_b$ and $r_d \approx r_{\rm max}$ & $t \lesssim t_{\rm b}$ and $r_d \ll r_{\rm max}$ &  $t_{\rm b} \ll t < t_{\rm mid} $ and  $r_{d} \ll r_{\rm max}$ &   $r_d \approx r_{\rm min}$ \\
\hline
$T_{\rm obs} \propto  $    &     $t^{-2/3}$     &    $ t^{(4-\alpha)/{(6\alpha+6)}}$     &       $ t^{(-3\alpha -2 )/ {(6 \alpha +6)}} $       &        $ t^{1/6} $  \\
\hline
$L_{\rm obs}  \propto $    &    $t^{-1/6}$       &     $t^{(4\alpha -4)/{(3\alpha +3)}}$   &       $ t^{- 16/{(3\alpha +3)} }$ &   $t^{-10/3} $   \\
\hline
\end{tabular}
\end{center}
\end{table*}

\begin{table*}
\begin{center}
\caption{Model parameters obtained for the three FBOTs from fitting the light curves and the color temperature evolution.}
\label{tab:my-table}
\begin{tabular}{|l|l|l|l|l|l|l|l|l|}
\hline
  &  $v_{\rm max}(c)$ & $v_b(c)$ & $v_{\rm min}(c)$ & $M_{\rm out}$($M_{\odot}$)  & $ t_{\rm end} (days) $ & $\alpha$ & $\eta$ \\
\hline
PS1-10bjp           & 0.13             & 0.05                   & 0.03 & 3.3 & 15 & 2.0& 1.5  \\
\hline
ZTF18abukavn            & 0.4              & 0.22                     & 0.11 & 4.3 & 13 & 3.0  & 1 \\
\hline
ATLAS19dqr         & 0.14       &  0.07           & 0.007    & 1.3 & 5.3 &4.3 & 2 \\
\hline
\end{tabular}
\end{center}
\end{table*}

\begin{figure}[h]
\includegraphics[scale=0.4]{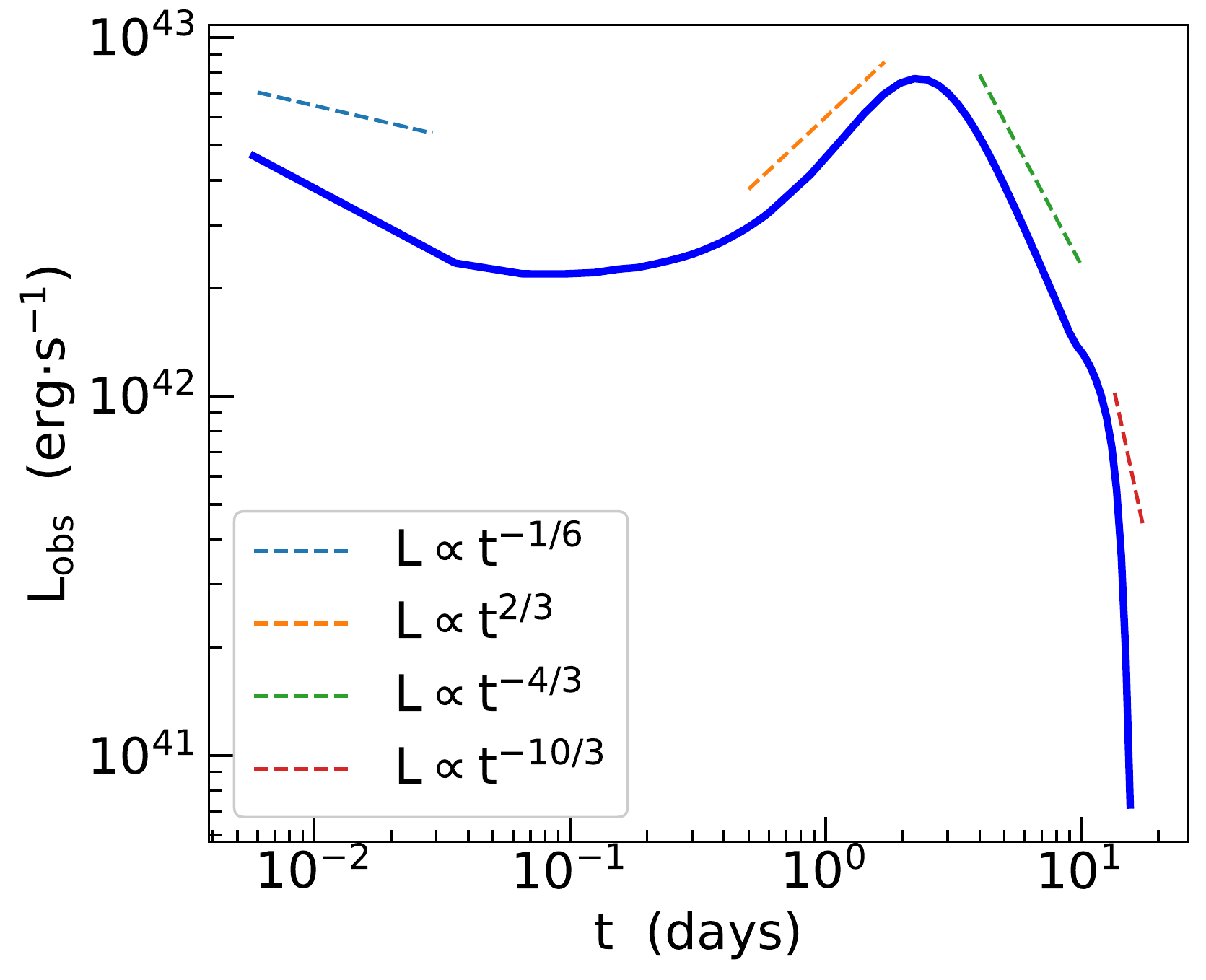}
\includegraphics[scale=0.4]{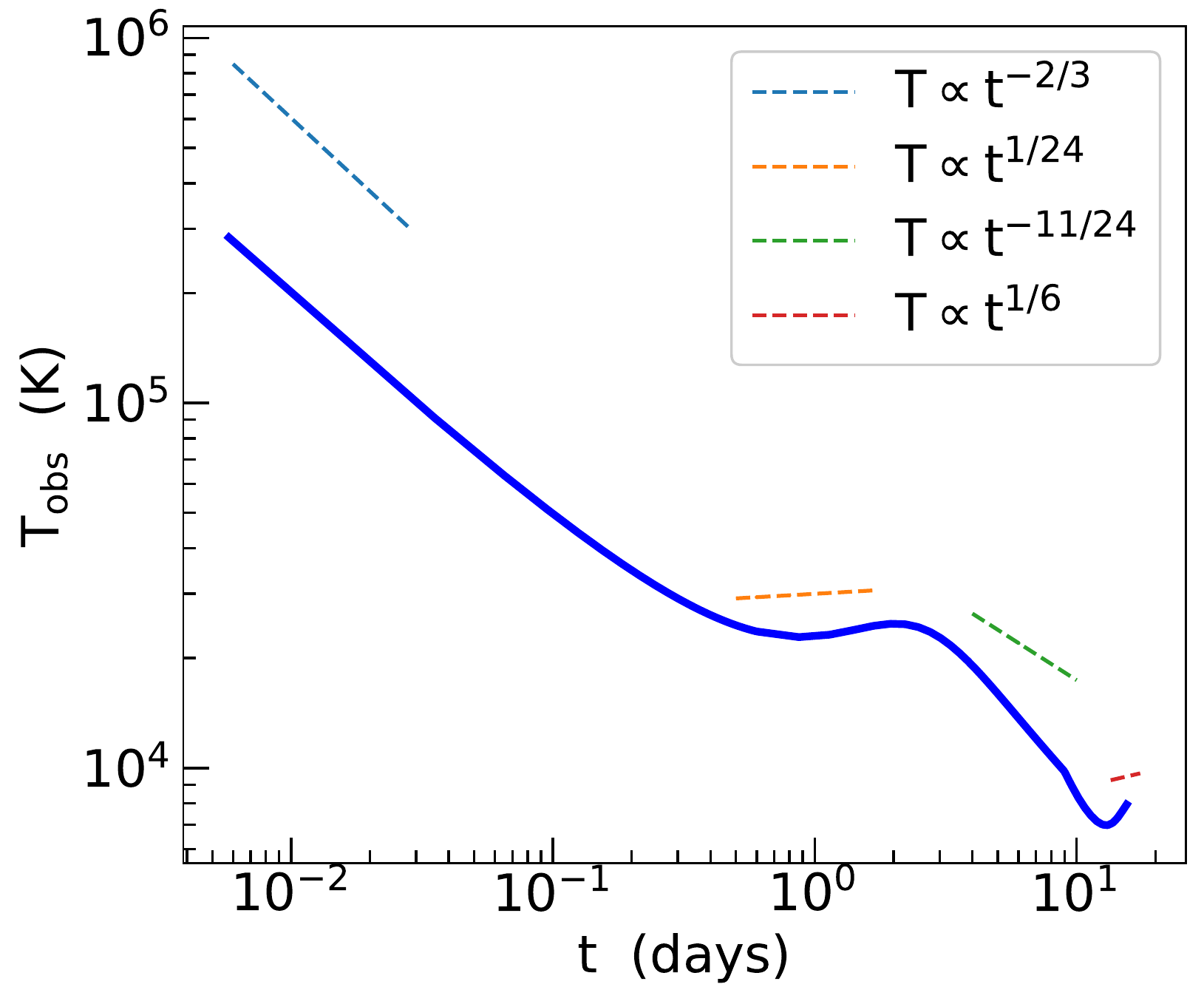}
\includegraphics[scale=0.4]{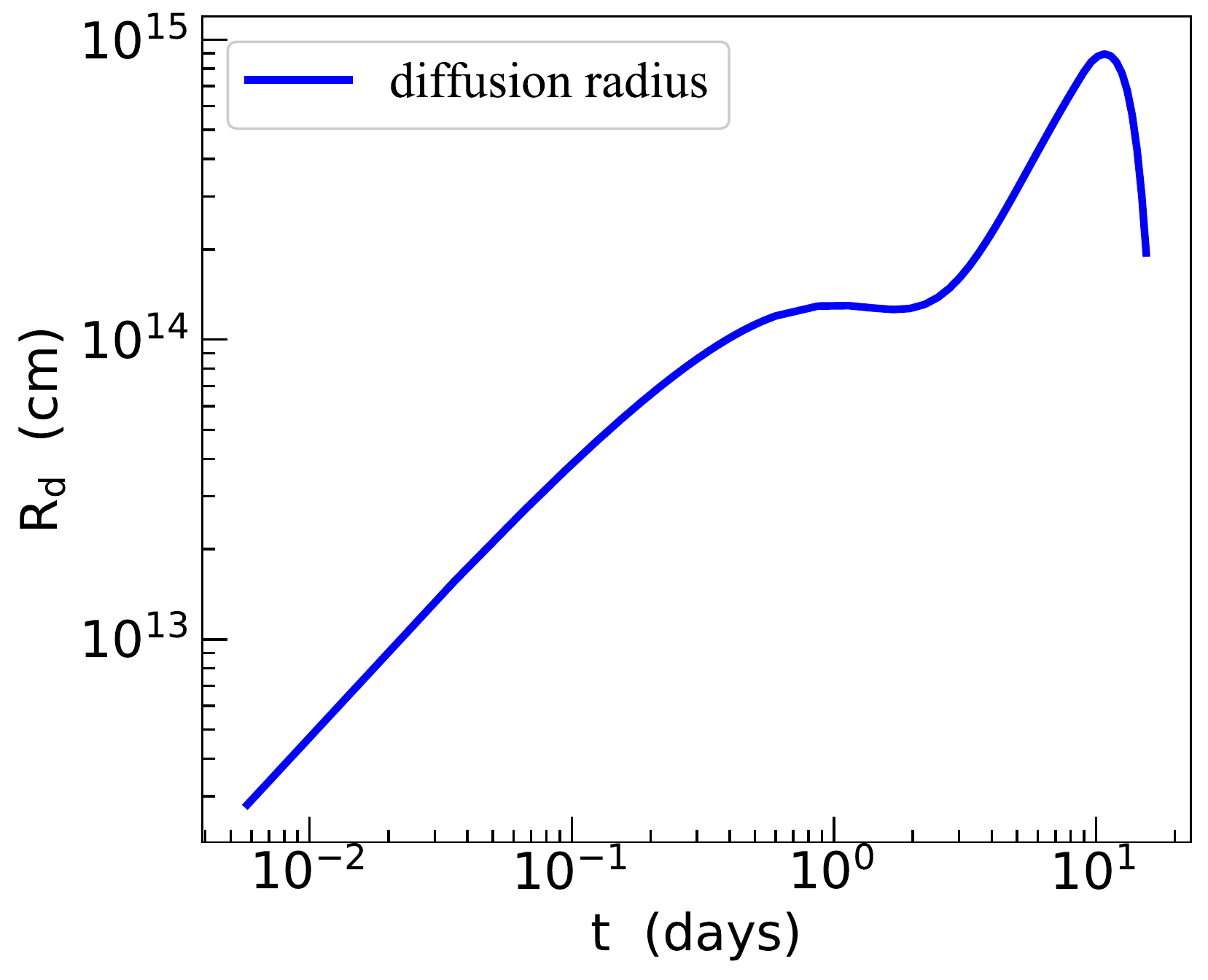}
\caption{Numerical results compare to analytical results.Top: light curve; middle: temperature evolves as time; bottom: diffusion radius evolves as time. Solid lines are our numerical results and dashed lines are analytical results. Parameter values are the same as in Figure \ref{dm_dt_t}.}
\label{Model result}
\end{figure}

%%%%%%%%%%%%%%%%%%%%%%%%
\subsubsection{Early times: $t \lesssim t_b$}
In this subsection, we analyze the asymptotic behavior of the diffusive emission in several stages of the outflow, starting from the earliest. Then we provide the numerical result for an example.

\begin{itemize}
\item Stage a): $r_d \approx r_{\rm max}$.
\end{itemize}

In this regime, the width of the velocity shell does not change significantly due to the difference in velocity, thus we could take $\rho \propto t^{-2}$ for the same velocity shell. And since $v_d \sim v_{\rm max}$, the diffusion shell is almost coincides with the shell with the maximum velocity, we could have $\rho_d \propto t^{-2}$. From Eq. \eqref{eq:T4}, we find $aT_{\rm obs}^4 \propto t^{-8/3}$ since $\rho_0$ is the initial density in the velocity shell with $v_d = v_{\rm max}$. Evaluating Eq. \eqref{eq:difshell} we find 
\beq
(r_{\rm max} -r_d)^2 \approx \frac{c t}{\kappa \rho_d}. 
\eeq
Taking the square root and the time derivative, we get 
\beq
v_{\rm max}- \frac{dr_d}{dt} \propto t^{1/2}.
\eeq

Therefore, from Eq. \eqref{eq:Ld} we see that $L_{\rm obs} \propto (v_d t)^2 aT^4_{\rm obs} (v_{\rm max}- \frac{dr_d}{dt}) \propto t^2 t^{-8/3} t^{1/2} = t^{-1/6}$. This stage is very short, only at the beginning of the ejection process.

%%%%%%%%%%%%%%%%%%%%%%%%
\begin{itemize}
\item Stage b): $r_d \ll r_{\rm max}$.
\end{itemize}

The width of velocity shell is dominated by the time difference during the shell ejection in this regime, which means the assumption $dr\sim vdt^{\prime}$ is reasonable.  For Eq. \eqref{eq:difshell}, since $r_d \ll r_{\rm max}$ and the ejected matter is concentrated in slow shells, we can write the approximate form as $\tau_v \sim \kappa \rho_d r_d \sim \kappa \rho_d \frac{dr}{dv} \cdot  v_d $. Therefore, according to Eq. \eqref{eq:difshell} we could consider $ \kappa \rho_d \frac{dr}{dv} \cdot  v_d = \frac{c}{v_d}$ when $r_d \ll r_{\rm max}$. Combine with Eqs. (\ref{rho},\ref{eq:tau}), we obtain $\tau_v \sim \kappa \frac{dm/dv}{4 \pi r_d^2} v_d \sim \frac{c}{v_d}$. We have $dm/dv \propto v_d^{-\alpha -1}$ at this stage according to Eq. \eqref{dm_dv}, thus we can write $\kappa \frac{dm/dv}{4 \pi r_d^2} v_d \sim \frac{v_d^{-\alpha -1}}{(v_d t)^2} v_d \sim \frac{c}{v_d}$ by taking $r_d \sim v_d t$, finally we could obtain $v_d \propto t^{-2/(\alpha +1)}$. Since we have obtained $v_d$, we can calculate $\rho_d \propto \frac{dm/dv}{r_d^2} \cdot \frac{1}{v_d} \propto t^{6/(\alpha+1)}$. Combine with Eqs. (\ref{rho}, \ref{eq:T4}), we can obtain  $aT^4_{\rm obs} \sim \rho_d v_d^2 (r_d)^{-2/3} \propto t^{(8-2\alpha)/(3\alpha +3)}$,  thus $T_{\rm obs} \propto t^{{(4-\alpha)}/{(6\alpha+6)}}$. Obviously we can work out $L_{\rm obs} \propto 4 \pi r^2_d aT^4_{\rm obs} v_d \propto t^{{(4\alpha -4)}/{(3\alpha +3)}}$ according to Eq. \eqref{eq:Ld} since we assume that $v_d - \frac{dr_d}{dt}$ is comparable to $v_d$ at this stage.

This result predicts a rising stage in the light curve for $\alpha > 1$. Due to the power law outflow ejection as Eq. \ref{dm_dv}, more outflow is constrained in lower velocity shells, which may carry more energy. As the diffusion shell recedes from the maximum velocity shell to the lower velocity shell, more energy will be released. Therefore, the luminosity will rise in this stage.

%%%%%%%%%%%%%%%%%%%%%%%%
\subsubsection{Middle times: $t_{\rm b} \ll t < t_{\rm mid} $}
\begin{itemize}
\item Stage c): $r_d \ll r_{\rm max}$.
\end{itemize}

The widths of velocity shells are dominated by the velocity difference in this regime, which means we have $dr\sim v(t-t^{\prime})$. Through similar procedure as in stage (b), we obtain $v_d \propto t^{-2/(\alpha +1)}$, and $\rho_d \sim \frac{dm/dv}{4\pi r_d^2} \propto t^{(3-\alpha)/(\alpha +1)} $. We can calculate $ aT^4_{\rm obs} \propto \rho_d v^2_d (\frac{\rho_d}{\rho_0})^{1/3} \propto \rho_d v^2_d (\frac{r^2_0 v_d}{r^2_d t})^{1/3} \propto t^{(-6 \alpha -4)(3 \alpha +3)}$, thus $T_{\rm obs} \propto t^{(-3\alpha -2 )/ {(6 \alpha +6)}} $ and $L_{\rm obs} \propto t^{- 16/{(3\alpha +3)} }$.

%%%%%%%%%%%%%%%%%%%%%%%%
\subsubsection{Late times: $t_{\rm mid} \ll t$}
\begin{itemize}
\item Stage d): $r_d \approx r_{\rm min}$.
\end{itemize}

At this stage, since the ejected matter is not concentrated in slow shells, we take the approximate form as $\tau_v \sim \kappa \rho_d (v_b - v_d) \cdot \frac{dr}{dv} \sim \kappa \rho_d v_b \cdot \frac{dr}{dv}$ due to $v_b \gg v_d$. We have $v_d \ll v_{\rm max}$, thus we consider $dm/dv$ to be a constant according to Eq (\ref{dm_dv}). At this stage, the density profile of the shell rapidly decreases as in Eq. \eqref{rho}, the diffusion radius will quickly move to the inner edge, that is, the velocity shell around the inner edge is just ejected and soon becomes the diffusion shell. Therefore, we will adopt this assumption $t - t^{\prime} \lesssim t_b$ for $\rho_d$ and $\rho$ at this stage. From Eq. \eqref{eq:difshell}, we can write $\tau_v \sim \kappa \frac{dm/dv}{(v_d t)^2} v_b \sim \frac{c}{v_d}$, therefore, $v_d \propto t^{-2}$, and $aT^4_{\rm obs} \propto \rho_d v^2_d (\frac{\rho_d}{\rho_0})^{1/3} \sim \frac{dm/dv}{r_d^2 v_d} v^2_d (\frac{r_0}{r_d})^{-2/3} \propto t^{2/3}$. Finally, we could obtain $T_{\rm obs} \propto t^{1/6} $ and $L_{\rm obs} \propto t^{-10/3}$.

The above asymptotic behavior is summarized in Table \ref{Analytical results}. Adopting $\alpha = 3 $ and $\gamma = 5/3$, a numerical example is shown in Figure \ref{Model result}. Clearly we can see the temperature slips from a few $\times 10^5 \ \rm K $ to $\sim 10^4 \ \rm K $ in the first one day, then it stands in a stable value of $\sim 10^4 \ \rm K $ and finally declines to a few $\times 10^3 \ \rm K $ slowly. For the light curve, the luminosity drops slowly when $t \ll 1 \ \rm day$, then it rises to the peak in the next a few days due to the increase of the outflow ejection rate. And after the peak the luminosity goes down rapidly, which coincides with the features of FBOTs. In our work, we assume that the ejection velocity of the outflow shell has a distribution as adopted in \cite{kashiyama15}, whereas in \cite{piro20}, the ejection velocity of the outflow shell is constant. This is the reason that leads to the difference between the results in \cite{piro20} and our results.

%%%%%%%%%%%%%%%%%%%%%%%%%%%%%%%
\section{Application} \label{section_4}

\begin{figure}
\includegraphics[scale=0.4]{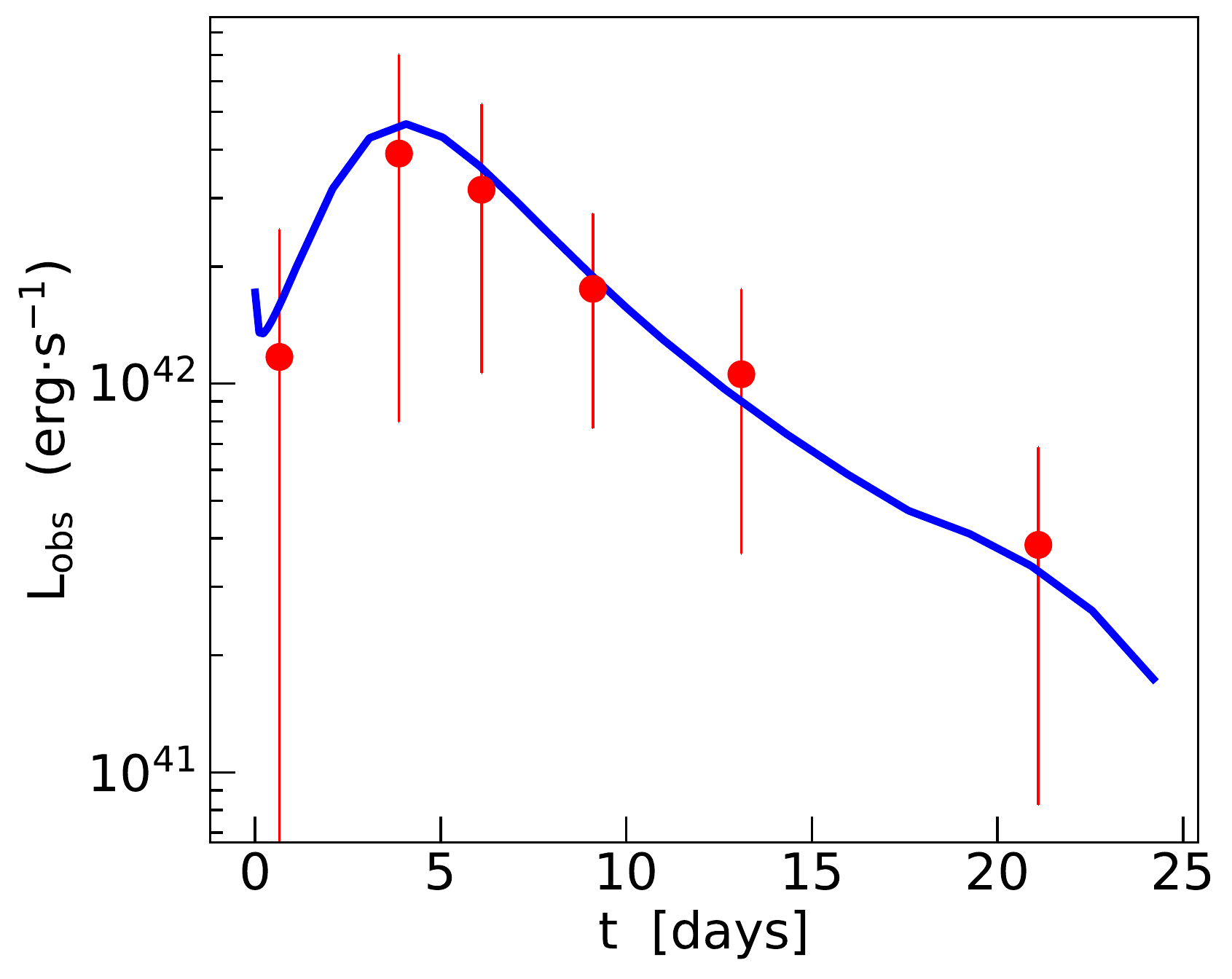}
\includegraphics[scale=0.4]{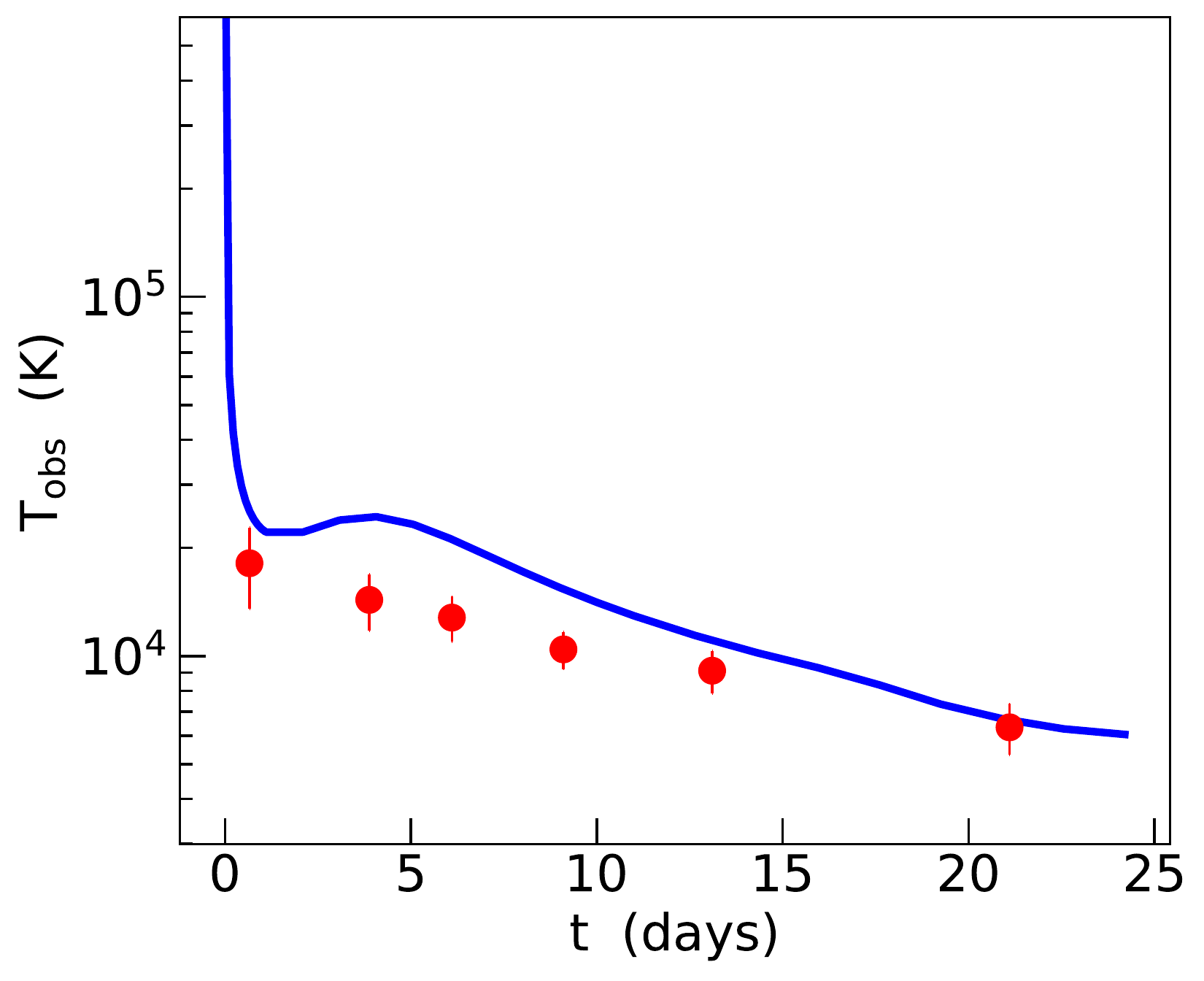}
\caption{Numerical results (blue solid line) compare to the observed data (red points) of PS1-10bjp. Top: light curve; bottom: temperature evolves as time. Data are from \cite{Drout2014}.}
\label{PS1-10bjp}
\end{figure}

\begin{figure}
\includegraphics[scale=0.4]{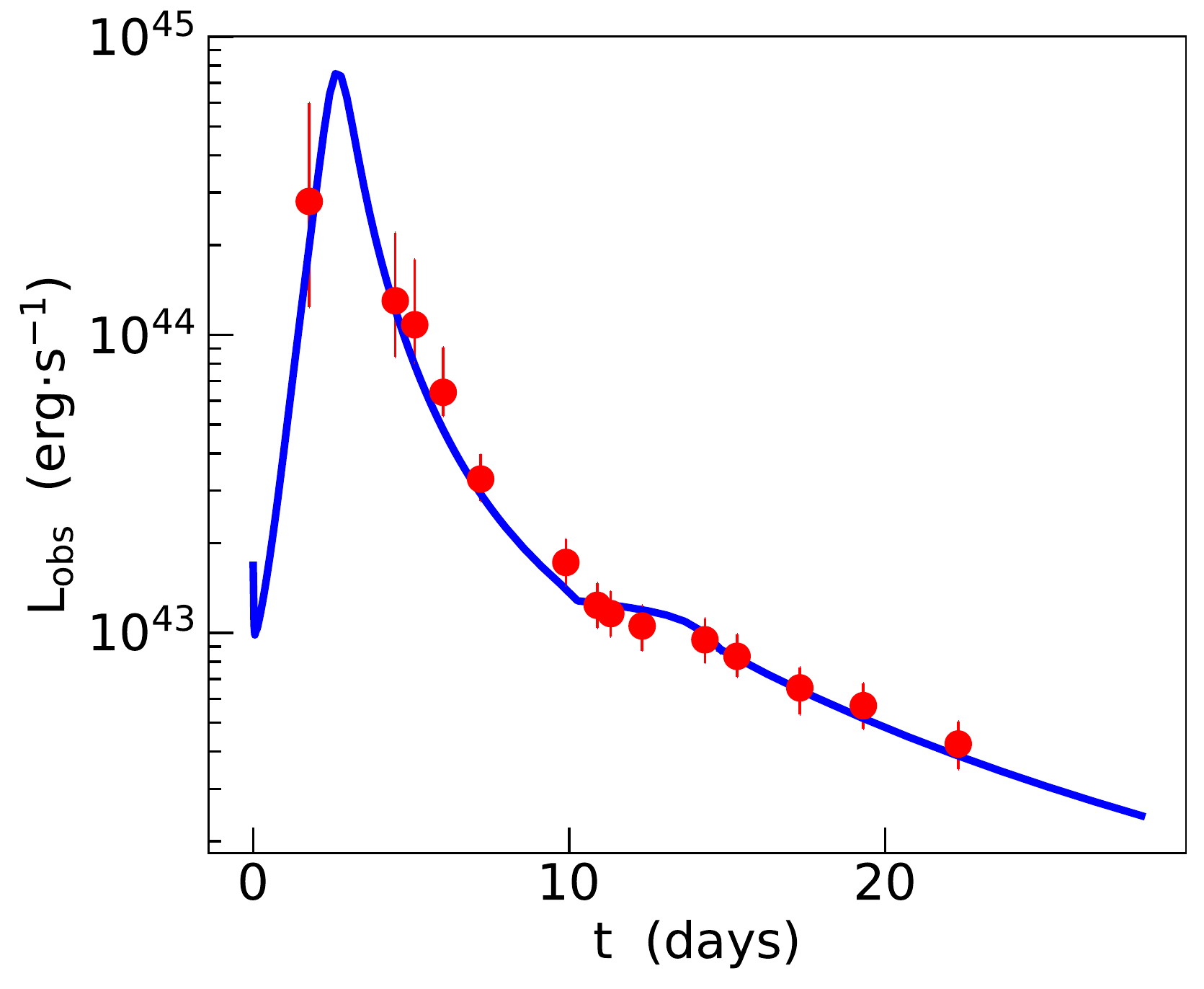}
\includegraphics[scale=0.4]{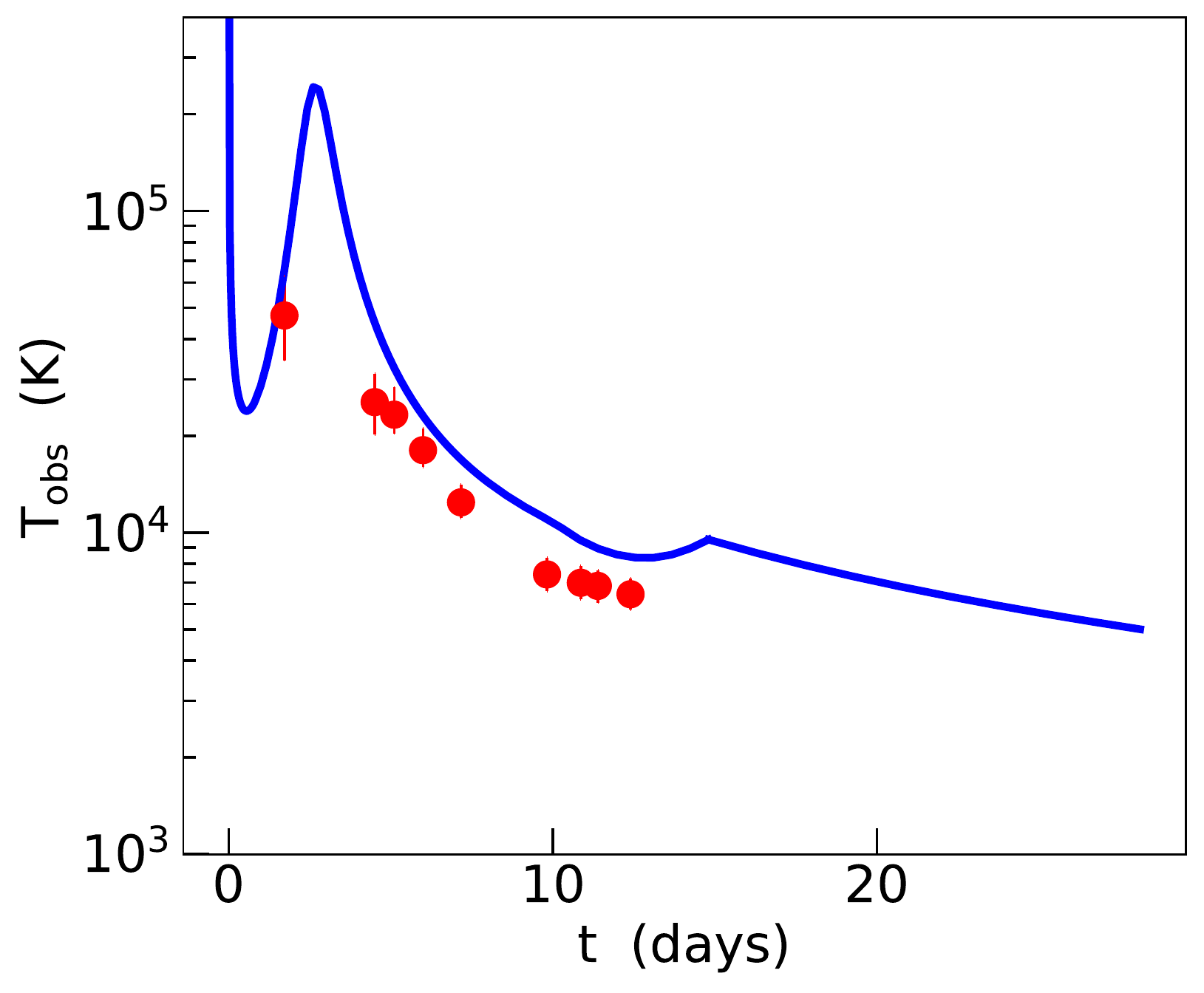}
\caption{Numerical results (blue solid line) compare to the observed data (red points) of ZTF18abukavn. Top: light curve; bottom: temperature evolves as time. Data are from \cite{Leung2021}.}
\label{ZTF18abukavn}
\end{figure}

\begin{figure}
\includegraphics[scale=0.4]{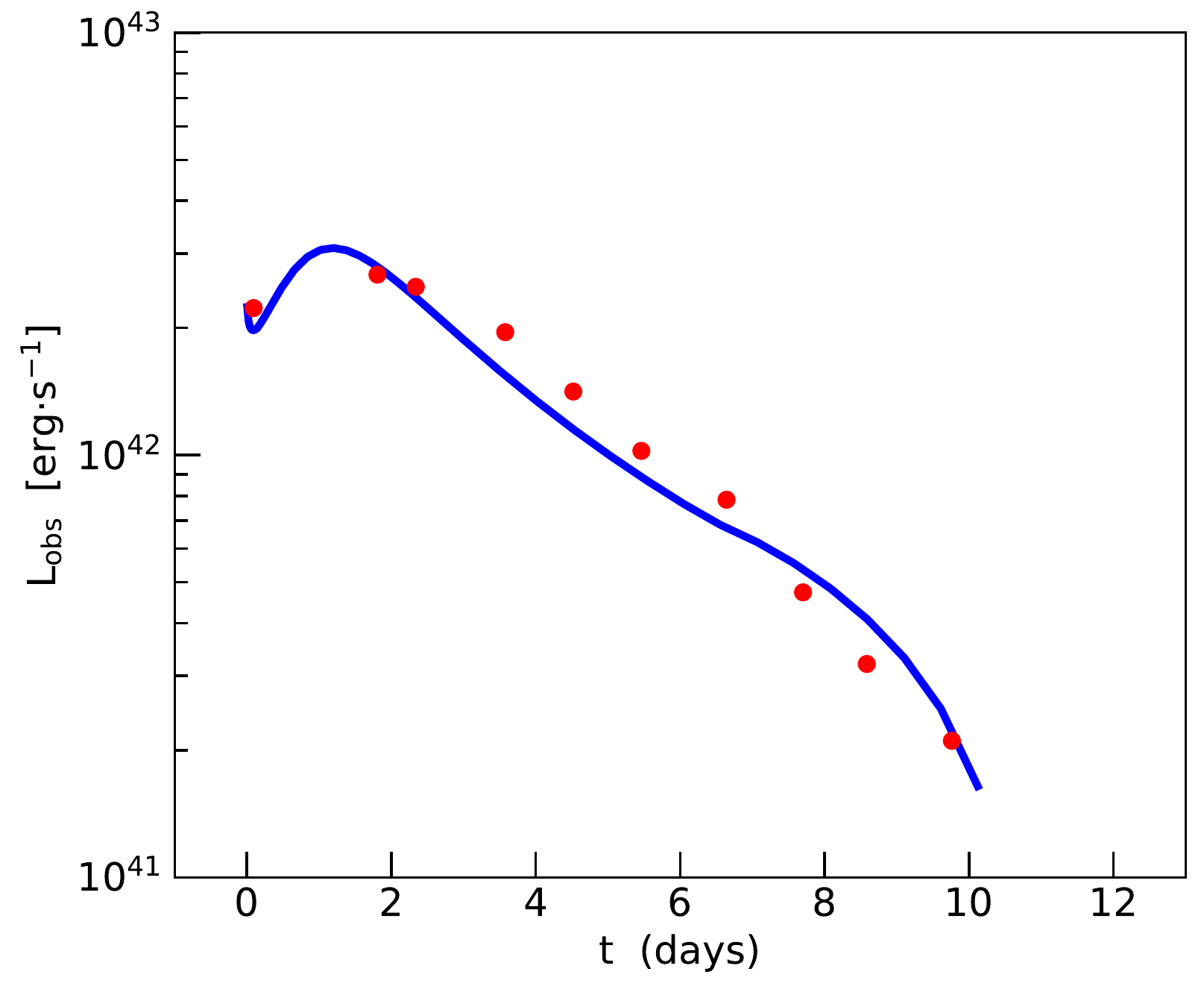}
\includegraphics[scale=0.4]{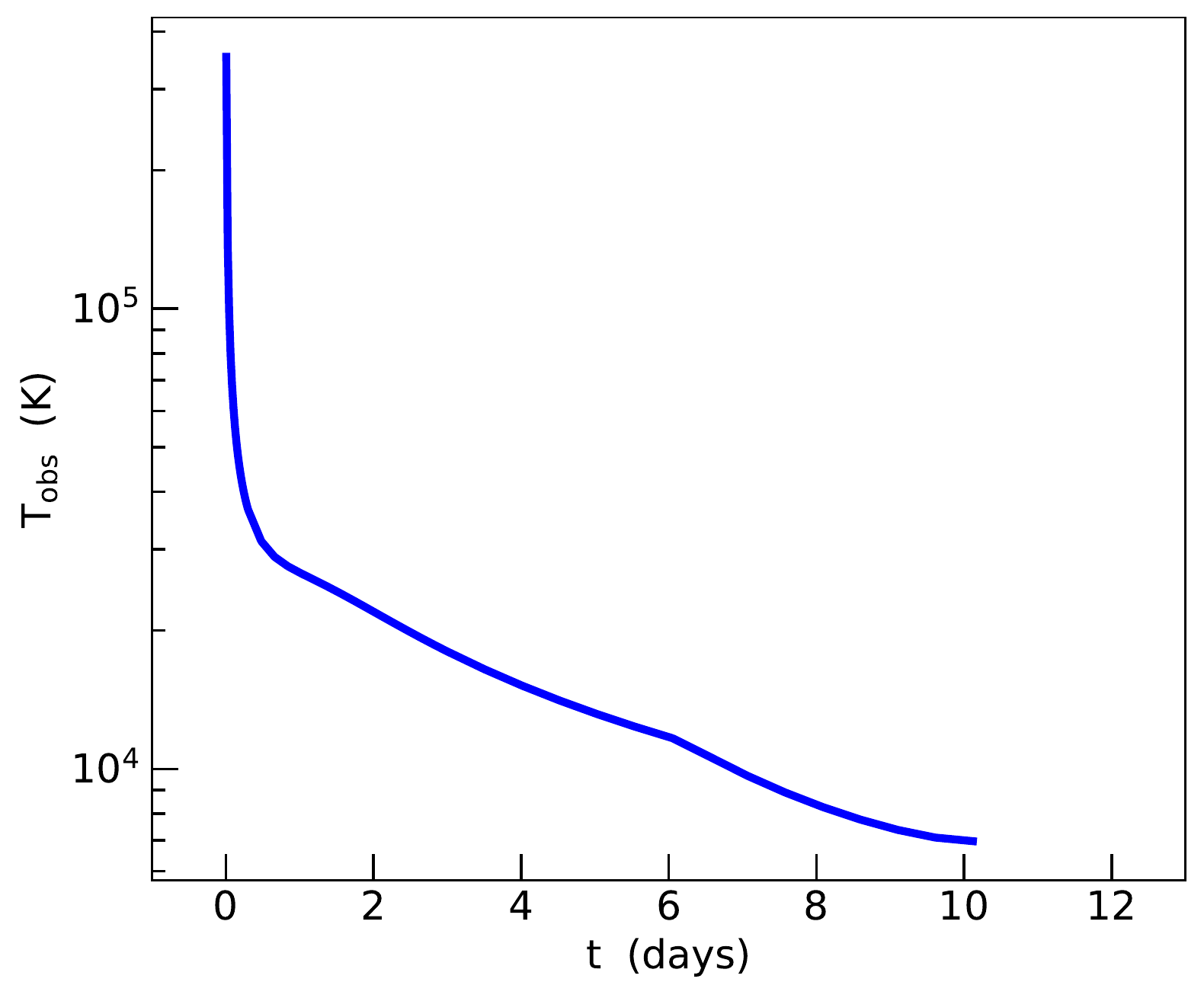}
\caption{Numerical results (blue solid line) compare to the observed data (red points) of ATLAS19dqr. Top: light curve; bottom: temperature evolves as time. Since the SED analysis is absent in \cite{chen2020}; \cite{Prentice2020}; \cite{zheng2021}, we plot only the model prediction for $T_{\rm obs}$. Data are from \cite{chen2020}.}
\label{ATLAS19dqr}
\end{figure}

Here we apply our model to three FBOTs. We put the well fit results of PS1-10bjp \citep[]{Drout2014} in Figure \ref{PS1-10bjp}, ZTF18abukavn in Figure \ref{ZTF18abukavn}, and ATLAS19dqr \citep[]{chen2020,zheng2021} in Figure \ref{ATLAS19dqr}. Fit parameters are presented in Table \ref{tab:my-table}.

Under the assumption of the black body, previous work has calculated the black body radius $r_{\rm BB}$ by $r_{\rm BB} \equiv (L_{\rm obs}/4 \pi a T_{\rm obs}^4 c)^{1/2}$, but it is not the radius where the photons escape. Since $L_{\rm obs}=4 \pi r_{\rm BB}^2 aT_{\rm obs}^4 c$ and we calculate $L_{\rm obs}$ via $L_{\rm obs}=4 \pi r_{d}^2 aT_{\rm obs}^4 c/\tau$ in our model, for a given $T_{\rm obs}$ and $L_{\rm obs}$, we always have $r_{d}>r_{\rm BB}$. Therefore, We do not compare $r_d$ with $r_{\rm BB}$, and we plot only the results of $T_{\rm obs}$ and $L_{\rm obs}$. Fit parameters are presented in Table \ref{tab:my-table}.

PS1-10bjp belongs to the golden sample of FBOTs in \cite{Drout2014}, and we compare our results to the data in Figure \ref{PS1-10bjp}. Initial work considers this transient as shock breakout or tidal disruption events in \cite{Drout2014}, but fails to obtain light curve fit about this event. In our model, luminosity declines during the initial 1 day. Since the lack of observed data, we are impossible to compare with the model. The next stage is  the rising curve which is consistent with our prediction that the bolometric light curve rises due to the accelerating mass outflow. And according to stage (b), since we adopt $\alpha = \rm 2.0$, obviously we could work out $L_{\rm obs} \propto t^{0.44}$. After the peak, luminosity declines as next two stages that all show in light cures. Easily we can see that $L_{\rm obs} \propto t^{-1.78}$ during the first decline stage according to stage (c). And in the second phase of decline, we can see that $L_{\rm obs} \propto t^{-10/3}$ which consists with stage (d). 

Though the light curve fits well, there are some discrepancies in the temperature curve fitting results, especially in the initial stage. Since we assume that the ratio of the internal energy to kinetic energy is constant as $\eta$ in Eq. \eqref{initial energy}, but in reality it may change over time. The model predicts that the temperature declines rapidly during the initial stage, which is inconsistent with the observed data. Then it rises briefly, and also no observed data supports this result.

ZTF18abukavn is also a typical FBOT,  and we compare our results with the data in Figure \ref{ZTF18abukavn}. Previous work interpreted this event as a result of  the ejecta interaction with the circumstellar matter, but lacked analytical forms for each stage \citep[]{Leung2021}. We describe analytical results as follow. At first, luminosity drops as stage (a), but no observed data in this stage. Then since we adopt $\alpha = \rm 3.0$, on the basis of stage (b), we obtain $L_{\rm obs} \propto t^{0.77}$. After the peak, light curve drops as $L_{\rm obs} \propto t^{-1.13}$ immediately, which corresponding to stage (c). Finally when $r_d =r_{\rm min}$, the diffusion radius coincides with the inner edge of the outflow. The outflow becomes transparent and experiences cooling in this stage. We adopt the conclusions from \cite{kashiyama15} that are $T_{\rm obs} \propto t^{-1}$ and $L_{\rm obs} \propto t^{-2}$ as a result of adiabatic cooling. Also, there are obvious errors in the temperature curve fitting results. 

ATLAS19dqr is another typical FBOT, for which we present our results in Figure \ref{ATLAS19dqr}. Previous work considered a central engine that keeps ejecting material but did not model the light curve well \citep[]{chen2020,Prentice2020,zheng2021}. As our results show, the luminosity drops initially, similar to the fore-mentioned two events. Then we obtain $L_{\rm obs} \propto t^{0.83}$ during initial $\sim \rm 2$ days when bolometric luminosity rises rapidly. Follow this stage, the bolometric luminosity declines as  $L_{\rm obs} \propto t^{-1.00}$ which couples with stage (c), and after this stage, luminosity drops as $L_{\rm obs} \propto t^{-10/3}$ which corresponds to stage (d). Since the SED analysis, thus the $T_{obs}$ data, is absent in \cite{chen2020}; \cite{Prentice2020}; \cite{zheng2021}, we plot only the model prediction for $T_{\rm obs}$.

%%%%%%%%%%%%%%%%%%%%%%%%%%

\section{Discussion} \label{section_5}

Given the obtained the parameters about the total outflow mass and the mass ejection timescale, we discuss the progenitors for FBOTs. First, for the stellar mass  larger than ${\rm 15 \rm{\ M_{\odot}}}$, at the end of the stellar evolution, the stellar core may collapse and form a black hole \citep[]{pejcha2015}. Therefore, the  mass of the outer envelop to be accreted larger than  $1 \rm {\ M_{\odot}}$ is reasonable.

Next, we discuss the outflow ejection timescale. The accretion disk mentioned in this paper is a thick accretion disk, which means that the material falling on the accretion disk can be immediately accreted. Therefore, the accretion timescale is equivalent to the free-fall timescale of the stellar outer layer material. Since the accretion timescale and the mass ejection timescale are also equivalent, we can estimate that the free-fall timescale of the outer layer of the star is about 10 days. By $t_{\rm free} =\pi (R_{*}^{3}/8GM_{\rm BH})^{1/2}$ \citep[]{kashiyama15}, where $t_{\rm free}$ is the free-fall timescale and $R_{*}$ is the radius of the stellar outer layer, for a stellar with a mass of 15 $M_{\odot}$ and an outer material free-fall timescale of 10 days, we can estimate $R_{*} \approx$ a few $10^{13}$ cm. Therefore, we consider the progenitor is a red giant with a mass of about 15 ${\ M_{\odot}}$ and a radius of about $10^{13}$ cm.

Our model results reveal a very early stage which the temperature shows a sharp drop. This temperature drop stage results from the rapid decline of the internal energy of the earliest shells at the diffusion radius due to the adiabatic cooling. At stage (a), the diffusion radius $r_d$ almost coincides with the outer edge, so the early drop of the temperature, thus the luminosity, is dominated by the decline of internal energy of the shells at this time. So far, no data is available at such an early time to support this result. With the future there may be telescopes with a larger field of view, we could obtain such early data. As more and more photons escape from the outflow, such energy drop becomes less and less dominant. 

We compare the kinetic energy of the outflow and the total radiative energy. Considering the kinetic energy as $E_k = \frac{1}{2} M_{\rm out} v^2$, since $M_{\rm out} \sim 1 M_{\odot}$ and $v \sim 0.1 c$, we can estimate the total kinetic energy as $\sim10^{52} \rm \ erg$. For the total radiative energy, we estimate it from the light curve, in which we can get a peak luminosity of $L_{\rm peak} \sim 10^{43}$ erg/s and the peak timescale $t_{\rm peak} \sim 5 $ day, which gives the total radiative energy $E_{\rm rad} \sim L_{\rm peak} t_{\rm peak} \sim 10^{48} \rm \ erg$. The latter is much lower than the kinetic energy, which means that only the minority of kinetic energy is converted into radiation. When the expanded outflow moves into the circumstellar matter (CSM) and interacts with the CSM, it may  release most of kinetic energy through a shock. We do not currently have observational evidence of the such radiation, possibly because the CSM is very thin, the ejected material does not fully release kinetic energy through the collision, thus we can not see such weak radiation. In fact, some FBOTs do have been observed such bright radio radiation in the hundreds of days after the explosion \citep[]{Coppejans2020}, which might be explained by our model.

%%%%%%%%%%%%%%%%%%%%%%%%%%%%%%%
\section{Conclusion} \label{section_6}  

In order to explain the observed features of FBOTs, including the fast evolution, high peak luminosity and blue color, we propose a radiative diffusion in a time-dependent outflow model. We assume such outflow is produced during the core-collapse of a massive star \citep[]{kashiyama15,Antoni2021}. However, since the outflow is optically thick at the beginning, the photons will be frozen in it initially. As the outflow expands, it gradually becomes optically thin, and more and more photons escape from it. We calculate the energy carried by the photons that escape in this way, and we can obtain the analytical form and numerical form of the light curve. 

We apply the model to three FBOTs, including PS1-10bjp, ZTF18abukavn, and ATLAS19dqr. From the results of data fitting, on the one hand, we require the total mass of the outflow material to be $\sim 1-5 M_{\odot}$, and we estimate that the mass of the progenitor is about 15 $M_{\odot}$ so that it is enough to produce such a large mass of the outflow. On the other hand, we require the mass ejection timescale to be 10 days, which suggests that the progenitor has a radius of $10^{13}$ cm,  i.e., it is a red supergiant.

%%%%%%%%%%%%%%%%%%%%%%%%%%%%%%%%
\section{Acknowledgements}
This work is supported by the China Manned Spaced Project (CMS-CSST-2021-B11), National Natural Science Foundation of China (12073091) and Guangdong Basic and Applied Basic Research Foundation (2019A1515011119).

\bibliographystyle{raa}

\label{lastpage}

\end{CJK*}
\end{document}